\documentclass[12pt]{article}
\usepackage{amssymb,amsmath,amsthm,amsfonts,amscd}
\usepackage{graphicx}
\date{\small}
\newcommand\be{\begin{equation}}
\newcommand\ee{\end{equation}}
\newcommand\bea{\begin{eqnarray}}
\newcommand\eea{\end{eqnarray}}

\newcommand{\fatalpha}{{\bf \alpha \kern -0.44em \alpha}}
\newcommand{\fatsigma}{{\bf \sigma \kern -0.54em \sigma}}
\newcommand{\tpchi}{{\bf D \kern -0.35em D}}
\newcommand{\llambda}{{\bf \lambda \kern -0.45em \lambda}}

\renewcommand{\theequation}{\arabic{equation}}
\renewcommand{\theequation}{\thesection.\arabic{equation}}
\bibliography{plain}
\title{\bf \large{ Nonlinear and linear entanglement witnesses for bipartite systems via exact convex optimization }} \vspace{20mm}
\author{ M. A. Jafarizadeh$^{a,b,c}$
\thanks{E-mail:jafarizadeh@tabrizu.ac.ir},
 A.Heshmati$^{a}$ \thanks{E-mail:heshmati@tabrizu.ac.ir},
 K. Aghayar$^{a}$ \thanks{E-mail:aghayar@tabrizu.ac.ir}
 \\
$^a${\small Department of Theoretical Physics and Astrophysics,
University of Tabriz, Tabriz 51664, Iran.}  \\ $^b${\small
Institute for Studies in Theoretical Physics and Mathematics,
Tehran 19395-1795, Iran.}\\$^c${\small Research Institute for
Fundamental Sciences, Tabriz 51664, Iran.}} \pagebreak
\pagebreak

\vspace{20mm}
\begin{document}
\maketitle \vspace{15mm}
\begin{abstract}
Linear and nonlinear entanglement witnesses for a given bipartite quantum systems are constructed. Using single particle feasible region, a way of constructing effective entanglement witnesses for bipartite systems is provided by exact convex optimization. Examples for some well known two qutrit quantum systems show these entanglement witnesses in most cases, provide necessary and sufficient conditions for separability of given bipartite system. Also this method is applied to a class of bipartite qudit quantum systems with details for d=3, 4 and 5.

\end{abstract}
\hspace{1cm}{\bf Keywords: }
non-linear and linear entanglement witnesses\\
\hspace{1cm}{\bf PACS number(s): 03.67.Mn, 03.65.Ud}
\section*{Introduction}
\par
Entanglement is one of the interesting features of quantum systems. It is used as a physical resource in realization of some important quantum information and quantum computation processes such as quantum parallelism \cite{Deutsch}, quantum cryptography \cite{Ekert}, quantum teleportation \cite{Bennett,Bouwmeester}, quantum dense coding \cite{Wiesner,Mattle}, reduction of communication complexity \cite{Cleve} and beating classical communication complexity bounds with entanglement \cite{HorodeckiE}. In these applications usually a source produces entangled particles and after receiving these particles to related parties, there is an important question for the parties - are these particles already entangled?
\par
One approach to distinguish entangled states is entanglement witness (EW) \cite{HorodeckiE,terhal}. A quantum state is entangled iff there exists a Hermitian operator W with $Tr(W \rho) < 0$ and $Tr(W \rho_{sep}) \geqslant 0$ for any separable state $\rho_{sep}$ \cite{Bruss}. In this case, the witness, W, detects the entanglement of the density matrix, $\rho$. Entanglement witnesses have some advantages with respect to other efficiently implementable witnesses such as concurrence \cite{Weinstein}. Recently there has been an increased interest in nonlinear EWs because of their improved detection with respect to linear EWs. A nonlinear EW is any bound on nonlinear function of observables which is satisfied by separable states but violated by some entangled states \cite{HorodeckiE}.
\par
Optimization problems occur in both classical and quantum physics \cite{Hartmann}. One of the important subclasses of these problems, is convex optimization where the related functions of the problems are convex. The importance of convex optimization is that in these optimizations, any locally optimal solution of the problem is guaranteed to be globally optimal \cite{Chong}. On the other hand, the set of all possible states of a quantum system that can occur in nature must be a convex set \cite{Bengtsson} and as the state  of a quantum system is fully characterized by density matrix of that system so the density matrix must be a convex function. Therefore, convex optimization is a natural optimization in quantum information.
\par
In this article we discuss the problems and advantages of the convex optimization approach for finding entanglement witnesses of a given system. The approach considered here, is similar to that of article \cite{222ConvexOptimization}, with some improvements such as feasible region for one particle, which this one enable us for detecting entanglement of bipartite systems in d dimension. In particular, it will be shown that for \emph{some} density matrices the whole entanglement region can be detected. This approach is essentially goal-directed but it may not work for all possible density matrices.

\par
Organized into five sections, the first section reviews the definition of an entanglement witness for a quantum system. The next section begins with the feasible region definition for a \emph{single} qudit followed by some feasible regions example for qudits. A technique of convex optimization algorithm for finding EWs is presented in section 3. Section 4 focuses on some examples for $3\times 3$ bipartite systems. Finally, section 5 deals with a class of $d\times d$ density matrices, including examples for entanglement detection conditions for $d=3$, $d=4$, and $d=5$.

\section{Entanglement Witness}
\par
A bipartite mixed state is called separable, if it can be prepared by two parties in a \emph{classical} way, that is, by agreeing over the phone on the local preparation of states. A density matrix that has been created in this way can only contain classical correlations. Mathematically this means: a mixed state $\rho$ is called separable iff it can be written as \cite{Bruss}
\begin{equation}\label{RoSep}
    \rho_{s}=\sum_{i} p_{i}|a_{i}\rangle \langle a_{i}|\otimes |b_{i}\rangle \langle b_{i}|
\end{equation}
otherwise it is \textit{entangled}. Here the coefficients $p_{i}$ are probabilities, i.e. $0 \leqslant p_{i} \leqslant1$ and $\sum_{i}p_{i}=1$.
\par
An entanglement witness W is a Hermitian operator such that $Tr(W \rho_{s}) \geqslant 0$, for all separable states $\rho_{s}$, and
there exists at least one entangled state $\rho_{e}$ which can be detected by W, that is $Tr(W \rho_{e}) < 0$. The existence of an EW for any entangled state is a direct consequence of Hahn-Banach theorem \cite{Hahn} and the fact that the space of separable density operators is convex and closed.
\par
For a bipartite $d\times d$ system, one can expand an EW as follows
\begin{equation}\label{bpco1}
    W=I_{d}\otimes I_{d}+\sum_{i=1}^{N_{1}}\sum_{j=1}^{N_{2}}A_{i,j} \hat{O_{i}}\otimes \hat{O'_{j}}
\end{equation}
where $I_{d}$ is a $d\times d$ identity matrix, $A_{i,j}$ are some parameters, and $O_{i}$s ( $O'_{j}$s ) are $N_{1}$ ( $N_{2}$ ) Hermitian operators from first ( second ) party Hilbert space.

\section{Single particle FR}
\par
Consider the following mapping
\begin{equation}\label{Mapping}
    \rho_{s}\mapsto Tr(W\rho_{s})
\end{equation}
This is a mapping from convex region of separable density matrices to a convex region called feasible region (FR). As a separable density matrix, $\rho_{s}$, can be written as a convex combination of product states (\ref{RoSep}), for one party ( say first one ) this map has terms like
\begin{equation}\label{SingleParticleMap}
    |a_{i}\rangle \langle a_{i}|\mapsto Tr(\hat{O_{i}}|a_{i}\rangle \langle a_{i}|)=P_{i}
\end{equation}
The set of $P_{i}$s defines a convex region for first particle and we will refer to this region as single particle FR.
\par
In this section we determine single particle FR. To this aim consider a single particle, a d-level quantum mechanical system, or a qudit. The possible states of this particle can be represented by state vectors in the associated d-dimensional Hilbert space. Each observable associated with this particle, is represented by Hermitian or self-adjoint linear operators acting on the corresponding Hilbert space. These operators can be expanded in terms of the traceless Hermitian matrices which are the matrix representations of $d^{2}-1$ independent infinitesimal generators of $SU(d)$ group ( the special unitary group which is a matrix Lie group of dimension $d^{2}-1$). So in the following we use these generators as EWs operator expansion basis. Specially, we consider two sets of operator expansions for a $d$ level single particle as follows.

\par
\textbf{First set}
\par

In this set we choose $d^{2}-1$ operators
\begin{equation}\label{1stSet}
    O_{i}=\sqrt{\frac{d}{2(d-1)}} \gamma_{i}
\end{equation}
where $\gamma_{i}$'s, the infinitesimal generators of $SU(d)$, are represented as traceless Hermitian matrices for a $d$ dimensional system.
The FR for this set of operators is defined by ( see Appendix A for a proof )
\begin{equation}\label{set1}
    \sum_{i=1}^{d^{2}-1} P_{i}^{2}\leqslant1
\end{equation}

\textbf{Second set}
\par
In this set we choose $d^{2}$ operators as follows. Every d-dimensional square matrix can be written in terms of square matrices $E_{_{ij}}$, which has the value 1 at the position $(i , j )$ and zeros elsewhere. Now one can define Hermitian traceless basis for d-dimensional matrices as follows \cite{Pf}. The off-diagonal basis are given by
$$O_{\alpha, \beta}=\frac{1}{\sqrt{2}}( E_{_{\alpha, \beta}}+E_{_{\beta,\alpha}} ) , \quad \alpha < \beta $$
$$O_{\alpha, \beta}=\frac{i}{\sqrt{2}}( E_{_{\alpha, \beta}}-E_{_{\beta,\alpha}} ) , \quad \alpha > \beta $$
and the diagonal basis are
$$
O_{\alpha,\alpha}=E_{\alpha,\alpha}
$$
where $\alpha,\beta=1,...,d$ and $E_{\alpha,\beta}=|\alpha\rangle\langle\beta|$.
For this set, the FR becomes ( Appendix A )
\begin{equation}\label{set3}
    \sum_{\alpha,\beta=1}^{d^{2}} |P_{\alpha,\beta}|^{2} \leqslant1.
\end{equation}

\section{Convex optimization formalism for bipartite systems}
\par
In this section we will express the construction of an EW for an arbitrary bipartite state in terms of a convex optimization. A convex optimization problem \cite{Boyd}, consists of optimizing an objective under some linear or nonlinear matrix equality and inequality constraints, precisely, we deal with a problem as follows
\hspace{-40mm}\begin{equation}\hspace{-30mm}\label{COP}
 \begin{array}{c}
  \mathrm{minimize} \    \ \hspace{10mm}f_{0}(x) \\
   \hspace{36mm}\mathrm{subject} \ \mathrm{to} \hspace{8mm}\ f_{i}(x)\leqslant 0$,   \hspace{5mm}$i=1,...,m. \\
   \hspace{63mm}h_{j}(x)= 0$,   \hspace{5mm}$j=1,...,p. \\
 \end{array}
\end{equation}
Here, the objective and the constraint functions are convex functions which means that they satisfy inequality $f_{i}(\alpha x +\beta y)\leqslant \alpha f_{i}(x)+\beta f_{i}(y)$, for all $x, y \in R$ and all $\alpha, \beta \in R$ with $\alpha +\beta = 1$, $\alpha \geqslant 0$, $\beta \geqslant 0$ and the equality constraint functions $h_{i}(x)= 0$ must be affine (A set $C\in\textbf{R}^{n}$ is affine if the line through any two distinct points in $C$ lies in $C$).
\par
We present a two-step convex optimization algorithm as follows. In the first step, recalling the definition of an EW, we impose the first condition, $Tr(W \rho_{s}) \geqslant 0$, for all separable states $\rho_{s}$. As a matter of fact, in this step $Tr(W \rho_{s})$ is the objective function and inside of the FR will be defined by the inequality constraints. In the second step, for a given $\rho$, we impose the second condition for an EW, $Tr(W \rho)< 0$. Now, the objective function ( which will be minimized ) is $Tr(W \rho)$, and the inequality constraints come from the solution of the first step. The following sub-sections provide more details on these two steps.

\subsection{First step, the nonnegativity conditions for $Tr(W\rho_{_{s}})$ }
Here the approach is similar to \cite{JAB} but for the convex optimization. Consider a non-positive Hermitian operator of the form in (\ref{bpco1}). We will attempt to choose the real parameters $A_{i,j}$ such that W becomes an EW. To this aim, we introduce the maps $P_{i} = Tr(O_{i} \rho_{s})$ and $P'_{j} = Tr(O'_{j} \rho_{s})$ for any separable state $\rho_{s}$. These maps, map the convex set of separable states into a bounded convex region which will be called feasible region. The first property of an EW is that its expectation value over any separable state is non-negative, i.e., the condition
$$
Tr(W\rho_{s}) = 1+\sum_{i,j}A_{i,j} P_{i}P'_{j}\geqslant 0
$$
must be satisfied for any point $(P_{1}, P_{2} , ...)$ of the feasible region. In order to satisfy this condition, it is sufficient that the minimum value of $Tr(W\rho_{s})$ be non-negative. Using standard convex optimization, we find this minimum value and then we impose the non-negativity condition on this minimum.
\par
For the bipartite system, as we studied before, the FRs are defined by $ \sum_{i=1}^{N_{1}} P_{i}^{2}\leqslant1 $ and $ \sum_{j=1}^{N_{2}} P_{j}'^{2}\leqslant1 $. These constraints in matrix notation become $P^{t} P\leqslant 1$ and $P'^{t} P'\leqslant 1$, where the vector $P$ is $P=(P_{_{1}},...,P_{_{d^{2}-1}})^{t}$. Then the problem can be written as a convex optimization problem in the form
$$
\mathrm{Minimize} \quad 1 + P^{t}A P'
$$
$$
\mathrm{subject \ to} \quad P^{t} P\leqslant 1
$$
$$
,\quad P'^{t} P'\leqslant 1.
$$
The Lagrangian associated with this problem, is given by
\begin{equation}\label{Lag1}
    L=1 + P^{t}A P'-\frac{1}{2}\lambda (P^{t} P -1)-\frac{1}{2}\lambda' (P'^{^{t}} P' -1)
\end{equation}
Any points satisfying \emph{Karush-Kuhn-Tucker}($KKT$) conditions, are primal and dual optimal, with zero duality gap. These conditions are\\
1. primal constraints: $P^{t} P-1\leqslant 0$ and $P'^{t} P'-1\leqslant 0$,\\
2. dual constraints: $\lambda\succeq 0$ and $\lambda' \succeq 0$,\\
3. complementary slackness: $\lambda (P^{t} P -1)=0$ and $\lambda' (P'^{^{t}} P' -1)=0$,\\
4. gradient of Lagrangian with respect to $P, P'$ vanishes: $\nabla L=0$.\\
Requiring these conditions for our problem, the forth condition $\nabla L=0$ reduces to
\begin{equation}\label{dl0}
    A P'=\lambda P
\end{equation}
and
\begin{equation}\label{dl1}
    P^{t}A=\lambda' P'^{^{t}}
\end{equation}
and the complementary slackness condition requires that $\lambda (P^{t} P -1)=0$ and $\lambda' (P'^{^{t}} P'-1)=0$, which reduce to $P^{t} P =1$ and $P'^{^{t}} P'=1$ noting that our problem is a minimization problem, so $\lambda>0$ and $\lambda'>0$. Multiplying (\ref{dl0}) by $P^{t}$ from left side and using the reduced complementary slackness condition, $P^{t} P =1$, yields
$$
P^{t}A P'=\lambda P^{t}P=\lambda
$$
also multiplication (\ref{dl1}) by $P'$ from right side and using $P'^{^{t}} P'=1$, leads to
$$
P^{t}A P'=\lambda' P'^{t}P'=\lambda'
$$
then $\lambda=\lambda'$. So, from (\ref{dl0}) and (\ref{dl1}) one can write
$$
A^{t}A P'=\lambda^{2} P'
$$
and
$$
A A^{t}P=\lambda^{2} P
$$
where, $\lambda^{2}_{i}$ for $i=1,...,d^{2}-1$, are eigenvalues of $A^{t}A$. If we choose $A$'s in a way that $\lambda^{2}_{i}\leqslant1$ (i.e. $-1\leqslant|\lambda_{i}|\leqslant1$) for all $i$'s, then the minimum of Lagrangian (\ref{Lag1}), $1+\lambda 1$, is nonnegative which leads to the nonnegativity of $Tr(W\rho_{_s})$.

\subsection{Second step, the detection condition for a given density amtrix}
For a given density matrix $\rho_{_g}$, the entanglement detection condition is the negativity of $Tr(W \rho_{_g})$ so we want to minimize this term. Again this problem can be written as a convex optimization problem in the following form
$$
\mathrm{Minimize} \quad Tr(W \rho_{_g})
$$
$$
s.t. \quad A^{t}A-I_{d}\leqslant 0
$$
As we will see, the solution to this problem yields a nonlinear EW. Lagrangian, $L$, associated with this problem is
$$
L=Tr(W \rho_{_g})+Tr[(A^{t}A-I_{d})Z]
$$
where $Z$ is a positive matrix which plays the role of the Lagrange multipliers associated with the inequality constraint $A^{t}A-I_{d}\leqslant 0$.

\par
Requiring the $KKT$ conditions for this problem, as in the previous subsection, yields the equations
$$
\tilde{\rho}+A(Z^{t}+Z)=0
$$
and
$$
A^{t}A=I_{d},
$$
where $\tilde{\rho}$ is a matrix with components
\begin{equation}\label{RoTilta}
    \tilde{\rho}_{i,j}=Tr(\rho_{_{g}} O_{i}\otimes O'_{j}).
\end{equation}
If $Z$ be a symmetric matrix (i.e., $Z^{t}=Z$) then
$\tilde{\rho}=-2AZ$ so that $\tilde{\rho}^{t} \tilde{\rho}=4 Z^{2}$ or
\begin{equation}\label{Z}
    Z=\frac{1}{2}(\tilde{\rho} ^{t}\tilde{\rho})^{\frac{1}{2}}
\end{equation}
and
\begin{equation}\label{A}
    A=-\frac{1}{2}\tilde{\rho} Z^{-1}.
\end{equation}
The matrix A, in general, has a nonlinear form which yields a nonlinear form for the associated EW.  Now the minimum of the Lagrangian becomes
\begin{equation}\label{MinLag}
    1-Tr[\sqrt{\tilde{\rho} ^{t}\tilde{\rho}} ]
\end{equation}
and the negativity of this term is the entanglement detection condition for the given density matrix.

\par
The remaining part of the paper gives examples for two-qutrit systems and higher dimensional systems. As we will see in these examples, this method mainly gives necessary and sufficient condition(s) for separability. However, this method can not detect entanglement for all bipartite systems. In \cite{Guhne}, the authors have provided a general method to improve the entanglement detection for bipartite systems via nonlinear expressions in EWs. Their approach is an iterative method and can yield useful approximation. However, there is only one example for two-qubits, and no example for higher dimensions.

\section{Examples for qutrit systems}
\par
\textbf{A. The two qutrit Horodecki states} \\
As first example illustrating the methodology, consider the state described in \cite{HoroRo1} given by
\begin{equation}\label{RoN2}
  \rho_{_{1}}=\frac{2}{7}|\psi_{+}\rangle \langle \psi_{+}| + \frac{\alpha}{7}\sigma_{+}+ \frac{5-\alpha}{7}\sigma_{-}
\end{equation}
with $0\leqslant \alpha\leqslant5$, $|\psi_{+}\rangle=\frac{1}{\sqrt{3}}\sum_{i=0}^{2}|ii\rangle$, $\sigma_{+}=\frac{1}{3}(|01\rangle\langle 01|+|12\rangle\langle 12|+|20\rangle\langle 20|)$ and $\sigma_{-}=\frac{1}{3}(|10\rangle\langle 10|+|21\rangle\langle 21|+|02\rangle\langle 02|)$.
This state is PPT for $1\leqslant \alpha \leqslant4$ and as it was shown in \cite{HoroRo1,Doherty1}, the state is bound entangled for $3< \alpha \leqslant4$, separable for $2\leqslant \alpha \leqslant3$ and free entangled for $4<\alpha\leqslant 5$. In \cite{Doherty1}, Doherty et al. introduced a non-decomposable EW which detects the Horodecki bound entangled states for $3< \alpha \leqslant4$.
\par
Here we work with the second set of operators introduced in the second section, the EW takes the form
$$
W=I_{3}\otimes I_{3}+\sum_{i,j=1}^{9}A_{i,j} \hat{O_{i}}\otimes \hat{O_{j}}
$$
and the $Z$ matrix in (\ref{Z}) is equal with
$$
Z=\frac{1}{126}
\left(
\tiny{
\begin{array}{ccccccccc}
 6 & 0 & 0 & 0 & 0 & 0 & 0 & 0 & 0 \\
 0 & 6 & 0 & 0 & 0 & 0 & 0 & 0 & 0 \\
 0 & 0 & 6 & 0 & 0 & 0 & 0 & 0 & 0 \\
 0 & 0 & 0 & 6 & 0 & 0 & 0 & 0 & 0 \\
 0 & 0 & 0 & 0 & 6 & 0 & 0 & 0 & 0 \\
 0 & 0 & 0 & 0 & 0 & 6 & 0 & 0 & 0 \\
 0 & 0 & 0 & 0 & 0 & 0 & 7+2c & 7-c  & 7-c \\
 0 & 0 & 0 & 0 & 0 & 0 & 7-c  & 7+2c & 7-c \\
 0 & 0 & 0 & 0 & 0 & 0 & 7-c  & 7-c  & 7+2c
\end{array}
}
\right)
$$
where, $c=\sqrt{19+3\alpha(\alpha-5)}$. The detection condition is given by
$$
1-\sqrt{19-15 \alpha+3 \alpha^2}<0
$$
which detects all of the entanglement region, i.e. for $0\leqslant \alpha<2$ and $3< \alpha\leqslant5$. So the necessary and sufficient condition for separability is obtained.

\textbf{B. Density matrix in unextendible product bases (UPBs)}
\par
The density matrix considered here, is constructed using UPBs \cite{UPBD} which is a bound entangled density matrix for $3\otimes 3$ state. The state has the following matrix form
\begin{equation}\label{33UPB}
\rho_{_{UPB}}=-\frac{1}{72}
\left(
\tiny{
  \begin{array}{ccccccccc}
    -7 & -7 & 2 & 2 & 2 & 2 & 2 & 2 & 2 \\
    -7 & -7 & 2 & 2 & 2 & 2 & 2 & 2 & 2 \\
    2 & 2 & -7 & 2 & 2 & -7 & 2 & 2 & 2 \\
    2 & 2 & 2 & -7 & 2 & 2 & -7 & 2 & 2 \\
    2 & 2 & 2 & 2 & -16 & 2 & 2 & 2 & 2 \\
    2 & 2 & -7 & 2 & 2 & -7 & 2 & 2 & 2 \\
    2 & 2 & 2 & -7 & 2 & 2 & -7 & 2 & 2 \\
    2 & 2 & 2 & 2 & 2 & 2 & 2 & -7 & -7 \\
    2 & 2 & 2 & 2 & 2 & 2 & 2 & -7 & -7 \\
  \end{array}
  }
\right)
\end{equation}
\par
Using the second set of FR relations, the detection is $Tr(W \rho_{_{UPB}})=-0.087$ and the matrix $A$ is
\begin{equation}\label{33UPBMatrixA}
A_{_{UPB}}=
\left(
\tiny{
  \begin{array}{ccccccccc}
    0.2909 & 0.1587 & 0.2909 & 0 & 0 & 0 & 0.1873 & 0.3316 & -0.8127 \\
    0.1587 & 0.8463 & 0.1587 & 0 & 0 & 0 & 0.3403 & -0.0416 & 0.3403 \\
    0.2909 & 0.1587 & 0.2909 & 0 & 0 & 0 & -0.8127 & 0.3316 & 0.1873 \\
    0 & 0 & 0 & 0 & 0 & 0 & 0 & 0 & 0 \\
    0 & 0 & 0 & 0 & 0 & 0 & 0 & 0 & 0 \\
    0 & 0 & 0 & 0 & 0 & 0 & 0 & 0 & 0 \\
    -0.8127 & 0.3403 & 0.1873 & 0 & 0 & 0 & -0.2868 & -0.1551 & -0.2868 \\
    0.3316 & -0.0416 & 0.336 & 0 & 0 & 0 & -0.1551 & -0.8545 & -0.1551 \\
    0.1873 & 0.3403 & -0.8127 & 0 & 0 & 0 & -0.2868 & -0.1551 & -0.2868 \\
  \end{array}
  }
\right).
\end{equation}
\par
\textbf{C. Horodecki state }
\par
In third example, we consider the $3\times 3$ bound entangled state introduced in Ref. \cite{Horodecki2} as
\begin{equation}\label{Horo2}
    \rho_{a}=\frac{1}{8a+1}
    \left(
    \tiny{
      \begin{array}{ccccccccc}
        a & 0 & 0 & 0 & a & 0 & 0 & 0 & a \\
        0 & a & 0 & 0 & 0 & 0 & 0 & 0 & 0 \\
        0 & 0 & a & 0 & 0 & 0 & 0 & 0 & 0 \\
        0 & 0 & 0 & a & 0 & 0 & 0 & 0 & 0 \\
        a & 0 & 0 & 0 & a & 0 & 0 & 0 & a \\
        0 & 0 & 0 & 0 & 0 & a & 0 & 0 & 0 \\
        0 & 0 & 0 & 0 & 0 & 0 & \frac{1+a}{2} & 0 & \frac{\sqrt{1-a^{2}}}{2} \\
        0 & 0 & 0 & 0 & 0 & 0 & 0 & a & 0 \\
        a & 0 & 0 & 0 & a & 0 & \frac{\sqrt{1-a^{2}}}{2} & 0 & \frac{1+a}{2} \\
      \end{array}
      }
    \right)
\end{equation}
This state is $PPT$ while it is entangled for all $0<a<1$.
Working with the second set of operators, we plot the detection in terms of parameter $a$, which shows that, all $PPT$ region
has been detected by our method ( see Figure 1 ).


\par
\section{Generalized qudit Choi maps}
In order to improve our knowledge of the convex optimization formalism for detecting entanglement of quantum systems, we will concentrate here on exploring $d\times d$ bipartite Choi maps density matrices \cite{GeneralizedChoi}. These density matrices have the following form
\begin{equation}\label{GChoiRo}
    \rho_{_{PPT}}=p | \psi_{0,0}  \rangle\langle  \psi_{0,0} |+\frac{1-p}{d}\sum_{i=1}^{d-1} \mu_{i} \rho_{i}, \quad p\leqslant \frac{1}{d}
\end{equation}
where $| \psi_{0,0}  \rangle=\frac{1}{\sqrt{d}}\sum_{l=0}^{d-1}|l\rangle|l\rangle$, $0\leqslant p\leqslant1$, $\rho_{i}=\sum_{l=0}^{d-1}|l\rangle\langle l|\otimes |l+i\rangle\langle l+i|, \quad i=0,1,...,d-1$ and $\mu_{i}$'s are parameters with properties $0\leqslant\mu_{i}\leqslant1, \quad \sum_{i=1}^{d-1}\mu_{i}=1$.
Also it has shown that ( the so called $PPT$ conditions )
\begin{equation}\label{GChoiRoPPTCondition}
    \rho^{T_{A}} \geqslant 0 \Rightarrow  p\leqslant min\{min\{\frac{\mu_{i}}{1+\mu_{i}}, \frac{\mu_{j} \mu_{k}}{\mu_{j} \mu_{k}-1}+\sqrt{\frac{\mu_{j} \mu_{k}}{(\mu_{j} \mu_{k}-1)^{2}}}   \}\}, \quad i\neq j\neq k=1,...,d-1.
\end{equation}
In the following we study the entanglement detection problem by the convex optimization formalism discussed previously and we give detection conditions in general.
\par
For density matrices (\ref{GChoiRo}), working with the second set of operators, we have
$$
\tilde{\rho}_{_{\alpha\beta,\mu\nu}}=Tr(\rho_{_{PPT}} O_{\alpha\beta}\otimes O_{\mu\nu}), \quad \alpha,\beta,\mu,\nu=0,...,d-1.
$$
The off diagonal elements of this matrix are
\begin{equation}\label{offdiag}
    \tilde{\rho}_{_{\alpha\alpha,\beta\beta}}=\frac{1-p}{d}\mu_{_{\beta-\alpha}}, \quad \alpha\neq\beta=0,...,d-1,
\end{equation}
where $\beta-\alpha$ is calculated in modulo $d$.
The diagonal elements are
\begin{equation}\label{diagdd}
    \tilde{\rho}_{_{\alpha\alpha,\alpha\alpha}}=\frac{p}{d}, \quad \alpha=0,...,d-1,
\end{equation}
also,
$$
\tilde{\rho}_{_{\alpha\beta,\alpha\beta}}=\frac{p}{d}, \quad \alpha<\beta,
$$
and finally for $\alpha>\beta$ we have
$$
\tilde{\rho}_{_{\alpha\beta,\alpha\beta}}=-\frac{p}{d}.
$$
The relations (\ref{offdiag}) and (\ref{diagdd}) form the following $d\times d$ block sub-matrix
$$
\tilde{\rho}_{_{d\times d}}=\frac{p}{d}I_{_{d}}+\frac{1-p}{d}\sum_{i=1}^{d-1}(\mu_{i}S^{i})=\frac{p}{d}I_{_{d}}+\frac{1-p}{d}V
$$
where $S$ is the shift operator for $d\otimes d$ defined as
$$
S=\left(
\tiny{
  \begin{array}{ccccc}
    0 & 1 & 0 & \ldots & 0 \\
    0 & 0 & 1 & \ldots & 0 \\
    \vdots & \vdots & \vdots & \ddots & \vdots \\
    1 & 0 & 0 & \ldots & 0 \\
  \end{array}
  }
\right)
$$
The eigenvalues of $(\tilde{\rho}^{t}\tilde{\rho})_{_{d\times d}}$ are
$$
\frac{1}{d}|p+(1-p)V_{k}|, \quad k=0,...,d-1
$$
where $V_{k}=\sum_{j=1}^{d-1}\mu_{j}\omega^{j k}$ and $\omega=e^{\frac{2 \pi i}{d}}$.
Finally the detection condition (\ref{MinLag}) for total density matrix, (\ref{GChoiRo}), is equal with
\begin{equation}\label{Detecdd}
    1-\left[ (d-1)p + \frac{1}{d}\sum_{k=0}^{d-1}|p+(1-p)V_{k}| \right ]< 0.
\end{equation}

\par
Now in the following we study the cases $d=3$ and $d=4$ in details, and we will obtain the necessary and sufficient conditions for separability in $d=3$.\\
\par
\textbf{A. The case d=3}\\
\par
It is instructive to examine our formalism for the case $d=3$ which leads to necessary and sufficient conditions for the entanglement detection. For $d=3$ explicitly we have
\begin{equation}\label{d3Choi}
    \rho_{_{PPT}}=p | \psi_{0,0}  \rangle\langle  \psi_{0,0} |+\frac{1-p}{3}(\mu_{1}\rho_{1}+\mu_{1}\rho_{1}), \quad p\leqslant \frac{1}{3}, \quad \mu_{1}+\mu_{2}=1
\end{equation}
where
$
| \psi_{0,0}  \rangle=\frac{1}{\sqrt{3}}(|0\rangle|0\rangle+|1\rangle|1\rangle+|2\rangle|2\rangle)
$
,
$\rho_{1}=|0\rangle \langle 0|\otimes |1\rangle \langle 1| +|1\rangle \langle 1|\otimes |2\rangle \langle 2|+|2\rangle \langle 2|\otimes |0\rangle \langle 0|
$
and
$
\rho_{2}=|0\rangle \langle 0|\otimes |1\rangle \langle 1| +|1\rangle \langle 1|\otimes |2\rangle \langle 2|+|2\rangle \langle 2|\otimes |0\rangle \langle 0|.
$
The $PPT$ condition for this density matrix is given by
$$
0\leqslant p\leqslant \frac{\mu_{1} \mu_{2}}{\mu_{1} \mu_{2}-1}+\sqrt{\frac{\mu_{1} \mu_{2}}{(\mu_{1} \mu_{2}-1)^{2}}}.
$$
Using the second set of operators, the detection condition for this $PPT$ density matrix (\ref{Detecdd}) is
$$
 {2-6p-[(1-3p)^{2}+3(p-1)^{2}(1-2\mu_{1})^{2}]^\frac{1}{2}} <0.
$$
Both $PPT$ and detection conditions are satisfied for the following regions
$$
0<\mu_{1}<\frac{1}{2}, \quad \frac{\mu_{1}}{1+\mu_{1}}<p\leqslant \frac{\mu_{1} \mu_{2}}{\mu_{1} \mu_{2}-1}+\sqrt{\frac{\mu_{1} \mu_{2}}{(\mu_{1} \mu_{2}-1)^{2}}}
$$
or
$$
\frac{1}{2}<\mu_{1} <1, \quad \frac{1-\mu_{1}}{2-\mu_{1}}<p\leqslant \frac{\mu_{1} \mu_{2}}{\mu_{1} \mu_{2}-1}+\sqrt{\frac{\mu_{1} \mu_{2}}{(\mu_{1} \mu_{2}-1)^{2}}}
$$
Other PPT regions, $0\leqslant p\leqslant\frac{\mu_{1}}{1+\mu_{1}}$ and $0\leqslant p\leqslant\frac{1-\mu_{1}}{2-\mu_{1}}$, which could not be detected, are correspond to separable regions ( for a proof see Appendix B ). Therefore applying our convex optimization formalism for density matrices (\ref{d3Choi}) yields the necessary and sufficient conditions for separability.\\
\par
\textbf{B. The case d=4}
\par
For $d=4$ the entanglement detection condition (\ref{Detecdd}) reduces to
\begin{equation}\label{Detec44}
    1-\left[ 3p + \frac{1}{4}\sum_{k=0}^{3}|p+(1-p)V_{k}| \right ]< 0.
\end{equation}
which is equal with
$$
1-[ 3p + \frac{1}{4}(1+2|p+(1-p)V_{1}|+|p+(1-p)V_{2}|) ]
$$
where we use $V_{0}=\mu_{1}+\mu_{2}+\mu_{3}=1$ and $V_{1}=\bar{V}_{3}$.
Simplifying this relation and using $\mu_{2}=1-\mu_{1}-\mu_{3}$ yields
$$
\frac{1}{4}\{ 3-12p-|p+(1-p)(1-2\mu_{1}-2\mu_{3})|
$$
\begin{equation}\label{44det1}
 -2 \sqrt{p^2-2p(1-p)(1-\mu_{1}-\mu_{3})+(1-p)^{2}[1-2\mu_1(1-\mu_1)-2\mu_3(1-\mu_3)]}\}<0.
\end{equation}
We can remove the absolute value signs and replace that terms with the following ones
\begin{equation}\label{Abs1}
    p+(1-p)(1-2\mu_{1}-2\mu_{3})
\end{equation}
and
\begin{equation}\label{Abs2}
    -[p+(1-p)(1-2\mu_{1}-2\mu_{3})]
\end{equation}
So, the detection condition (\ref{44det1}) with $PPT$ condition (\ref{GChoiRoPPTCondition}), define two possible detection regions. These regions are presented in table1 and table2 of the appendix C, which are correspond to (\ref{Abs1}) and (\ref{Abs2}) respectively.
In Figure 2. these regions are plotted.
\\
\par
\textbf{C. The case $d=5$}
\par
For $d=5$ the detection condition (\ref{Detecdd}) is equal with
\begin{equation}\label{d=5Detection}
    1-\{ 4p+\frac{1}{5}[ 1+2(|p+(1-p)V_{1}|+|p+(1-p)V_{2}|)]\}<0.
\end{equation}
As a special case we assume $Im V_{1}=0$ which gives $\mu_{1}=\frac{1}{2}[(1-\sqrt{5})\mu_{2}+(\sqrt{5}-1)\mu_{3}+2\mu_{4}]$ and $\mu_{2}=\frac{\mu_{3}(1+\sqrt{5})+2(2\mu_{4}-1)}{\sqrt{5}-3}$ where we use $\sum_{i=1}^{4}\mu_{i}=1$. The detected PPT region under these simplifying assumptions plus PPT conditions (\ref{GChoiRoPPTCondition}) for $d=5$, is plotted in Figure 3.

\par
\section*{Summary and conclusion}
In this paper we have defined single particle feasible region ( FR ) and have used it for constructing entanglement witnesses for mixed bipartite systems by the convex optimization method. The advantage of considering the single particle FR is that it produces a straightforward formulation in the convex optimization method. We have also given an explicit method to construct nonlinear EWs for given bipartite systems. The ability of this approach is illustrated by applying it to the separability problem for \textit{some} well known two qutrit systems. However, this method can not detect entanglement for every bipartite systems.
\par
This approach leaves some questions open such as the optimality of these EWs. So a natural next step would be to investigate the optimality of the obtained EWs. Is there any relationship between these EWs ( specially, those EWs leading to necessary and sufficient conditions for separability ) and the optimal entanglement witnesses condition considered in \cite{Bertlmann} ? However, such problems which are under investigation, perhaps will probe some aspects of optimal EWs problems.

\newpage
\vspace{1cm}\setcounter{section}{0}
\setcounter{equation}{0}
\renewcommand{\theequation}{A-\roman{equation}}
  {\Large{Appendix A }}\\
\textbf{  a). Proving (\ref{set1})}
\par
First we prove that working in any basis, will not change the quantity $\sum_{i=1}^{d^{2}-1} P_{i}^{2}$.
Recalling $P_{i}$ mapping definition, $P_{i}=Tr( O_{i}|\alpha \rangle\langle\alpha|)$, we have
$$
\sum_{i=1}^{d^{2}-1} P_{i}^{2}=\sum_{i=1}^{d^{2}-1} \langle \alpha |O_{i}| \alpha \rangle \langle \alpha |O_{i}| \alpha \rangle
$$
If we change the basis $| \alpha \rangle=U | \beta \rangle$ where $U$ is a $d\times d$ unitary operator, then
$$
\sum_{i=1}^{d^{2}-1} P_{i}^{2}=\sum_{i=1}^{d^{2}-1}  \langle \beta |U^{\dag} O_{i}U| \beta\rangle \langle \beta |U^{\dag} O_{i}U| \beta\rangle
$$
Using the adjoint representation of the Lie algebra,
$$
U^{\dag} O_{i}U=\sum_{j=1}^{d^{2}-1}A_{i,j} O_{i}
$$
where $O_{i}$'s are generators of the Lie group, and $A$'s are a subgroup of $d\times d$ orthogonal matrices. In fact one can expand the generators of the Lie algebra as above. So we have
$$
\sum_{i=1}^{d^{2}-1} P_{i}^{2}=\sum_{j,k=1}^{d^{2}-1}( \sum_{i=1}^{d^{2}-1}A_{i,j}A_{i,k} )\langle \beta |O_{j}| \beta \rangle \langle \beta |O_{k}| \beta \rangle=\sum_{i=1}^{d^{2}-1}|\langle \beta |O_{i}| \beta \rangle |^{2}
$$
( the term in parenthesis,  $\sum_{i=1}^{d^{2}-1}A_{i,j}A_{i,k}$, is equal to $\delta_{j,k}$) which completes the proof.
\par
Now we choose $|\beta \rangle=[1,0,0,...,0]^{T} $, a $d-1$ dimensional vector and a appropriate set of generators, ${\gamma_{i}}$, \cite{Hioe},
$$
\hat{u}_{j,k}\equiv \hat{P}_{j,k}+\hat{P}_{k,j}
$$
$$
\hat{v}_{j,k}\equiv -i (\hat{P}_{j,k}-\hat{P}_{k,j})
$$
\begin{equation}\label{HioeSetW}
    \hat{w}_{l}\equiv -\sqrt{\frac{2}{l(l+1)}}[(\sum_{k=1}^{l}\hat{P}_{k,k})-l \hat{P}_{l+1,l+1}]
\end{equation}
where $\hat{P}_{m,n}\equiv|m\rangle\langle n|$, $1\leqslant j< k \leqslant d$ and $1\leqslant l \leqslant d-1$.
All of these operators are traceless and they generate the algebra $su(d)$. With these choices, we obtain
$$
\sum_{i=1}^{d^{2}-1} P_{i}^{2}=\sum_{i=1}^{d^{2}-1}|\langle \beta |O_{i}| \beta \rangle |^{2}=(\frac{d}{2(d-1)})\sum_{i=1}^{d^{2}-1} |\langle \beta |\gamma_{i}| \beta \rangle |^{2}
$$
All of the operators in (\ref{HioeSetW}) vanish in summation unless $\hat{w}_{_{l}}$s, which for them we have
$$
|\langle \beta |\hat{w}_{l}| \beta \rangle |^{2}=\frac{2}{l(l+1)}
$$
and as the number of such terms is $d-1$ so
$$
(\frac{d}{2(d-1)})\sum_{i=1}^{d^{2}-1} |\langle \beta |\gamma_{i}| \beta \rangle |^{2}=
(\frac{d}{2(d-1)}) \sum_{k=1}^{d-1} \frac{2}{l(l+1)}
$$
$$
=(\frac{d}{2(d-1)}) \sum_{l=1}^{d-1}(\frac{2}{l}-\frac{2}{l+1})=(\frac{d}{2(d-1)})(2-\frac{2}{d})=1
$$
As we interest to the entire FR, then
$$
\sum_{i=1}^{d^{2}-1} P_{i}^{2}\leqslant 1.
$$

\newpage
\vspace{1cm}\setcounter{section}{0}
\setcounter{equation}{0}
\renewcommand{\theequation}{A-\roman{equation}}
  {\Large{Appendix B}}\\
\textbf{  a). Proving separability}
\par
Here we prove that for PPT regions $0\leqslant p\leqslant\frac{\mu_{1}}{1+\mu_{1}}$ and $0\leqslant p\leqslant\frac{1-\mu_{1}}{2-\mu_{1}}$, the density matrix (\ref{d3Choi}) is separable.
\par
For first region, $0\leqslant p\leqslant\frac{\mu_{1}}{1+\mu_{1}}$, if $p=0$ then $\rho_{_{PPT}}=\frac{1}{3}(\mu_{1}\rho_{1}+\mu_{1}\rho_{1})$ which is obviously separable and if $p=\frac{\mu_{1}}{1+\mu_{1}}$ then $\rho_{_{PPT}}$ could be written as $\rho_{_{PPT}}=\frac{1}{1+\mu_{1}}\{3\mu_{1}[\frac{1}{3}(| \psi_{0,0}  \rangle\langle  \psi_{0,0} |+\frac{\rho_{1}}{3}+\frac{\rho_{2}}{3})]+(1-2\mu_{1})\frac{\rho_{2}}{3}\}$. The term $[\frac{1}{3}(| \psi_{0,0}  \rangle\langle  \psi_{0,0} |+\frac{\rho_{1}}{3}+\frac{\rho_{2}}{3})]$ has been explicitly represented as a mixture of product states \cite{HoroSep} so the later $\rho_{_{PPT}}$ is separable.
\par
For second region the separability proof is similar to the first region.

\newpage
\vspace{1cm}\setcounter{section}{0}
\setcounter{equation}{0}
\renewcommand{\theequation}{A-\roman{equation}}
  {\Large{Appendix C}}\\
\textbf{Detected entangled regions for generalized qudit Choi maps, $d=4$}\\
\par
Table 1.\\
\begin{tabular}{|c|c|c|}
  \hline
  $\mu_{3}$ & $\mu_{1}$ & $p$ \\\hline\hline

                               & $b_{1}<\mu_{1}\leqslant b_{2}$ & $c_{1}<p\leqslant c_{2}$ \\\cline{2-3}
  $a_{1}<\mu_{3}<\frac{4}{33}$ & $b_{2}<\mu_{1}< b_{3}$ & $c_{3}\leqslant p\leqslant c_{2}$ \\\cline{2-3}
                               & $\mu_{1}=b_{3}$ & $p=c_{3}$ \\\hline\hline

  $\frac{4}{33}\leqslant\mu_{3}\leqslant\frac{1}{4}$ & $b_{1}<\mu_{1}<b_{2}$ & $c_{1}<p\leqslant c_{2}$ \\\cline{2-3}
                                                     & $b_{2}<\mu_{1}<b_{3}$ & $c_{3}\leqslant p\leqslant c_{2}$ \\\hline\hline

                                    & $b_{1}<\mu_{1}< \mu_{3}$ & $c_{1}<p\leqslant c_{2}$ \\\cline{2-3}
                                    & $\mu_{3}<\mu_{1}<b_{2}$ & $c_{1}<p\leqslant c_{2}$ \\\cline{2-3}
  $\frac{1}{4}<\mu_{3}<\frac{1}{3}$ & $\mu_{1}=b_{2}         $ & $c_{3}<p\leqslant c_{2}$ \\\cline{2-3}
                                    & $b_{2}<\mu_{1}<b_{3}   $ & $c_{3}\leqslant p\leqslant c_{2}$ \\\cline{2-3}
                                    & $\mu_{1}=b_{3}$ & $p=c_{3}$ \\\hline\hline

  $\mu_{3}=\frac{1}{3}$ & $b_{4}<\mu_{1}<\frac{1}{3}$ & $c_{4}<p\leqslant c_{5}$ \\\cline{2-3}\hline

                                    & $b_{1}<\mu_{1}< b_{2} $ & $c_{1}<p\leqslant c_{2}$ \\\cline{2-3}
  $\frac{1}{4}<\mu_{3}<\frac{1}{3}$ & $\mu_{1}=b_{2}        $ & $c_{3}<p\leqslant c_{2}$ \\\cline{2-3}
                                    & $b_{2}<\mu_{1}<b_{3}  $ & $c_{3}\leqslant p\leqslant c_{2}$ \\\cline{2-3}
                                    & $\mu_{1}=b_{3}        $ & $p=c_{3}$ \\\hline\hline

  $\frac{4}{33}\leqslant\mu_{3}\leqslant\frac{1}{4}$ & $\mu_{1}=b_{2}$ & $c_{3}<p\leqslant c_{2}$ \\\cline{2-3}
                                                     & $\mu_{1}=b_{3}$ & $p= c_{3}$ \\ \hline

\end{tabular}\\
\par
where
\par
$a_{1}=\frac{1}{29}(9-4\sqrt{2})$,
$a_{2}=\frac{1}{29}(9+4\sqrt{2})$,

$b_{1}=4+97\mu_{3}-28\sqrt{\mu_{3}+12\mu_{3}^{2}}$,
$b_{2}=\frac{1}{71}[48-73\mu_{3}-4\sqrt{2}\sqrt{ (3\mu_{3}-1)^{2}}]$,

$b_{3}=\frac{1}{8}(4-7\mu_{3}+\sqrt{\mu_{3}(8-15\mu_{3})})$,
$b_{4}=\frac{1}{3}(109-28\sqrt{15})$,

$$
c_{1}=1+\frac{28+6(\mu_{1}+\mu_{3})}{(\mu_{1}-\mu_{3})^{2}-16(\mu_{1}+\mu_{3})-32}-2\sqrt{\frac{15(\mu_{1}^{2}+\mu_{3}^{2})-12(\mu_{1}+\mu_{3})+6\mu_{1}\mu_{3}+4}{[(\mu_{1}-\mu_{3})^{2}-16(\mu_{1}+\mu_{3})-32]^{2}}}
$$

$c_{2}=\sqrt{\frac{\mu_{1}\mu_{3}}{(\mu_{1}\mu_{3}-1)^{2}}}+\frac{\mu_{1}\mu_{3}}{\mu_{1}\mu_{3}-1}$,
$c_{3}=\frac{2(\mu_{1}+\mu_{3})-1}{2(\mu_{1}+\mu_{3})}$,
$c_{4}=\frac{3\mu_{1}(3\mu_{1}-6\sqrt{15}-32)+6\sqrt{15}-65}{3\mu_{1}(3\mu_{1}-50)-335}$, $c_{5}=\frac{\sqrt{\mu_{1}}}{\sqrt{\mu_{1}}+\sqrt{3}}$.


Table 2.
\par
\begin{tabular}{|c|c|c|}
  \hline
  $\mu_{3}$ & $\mu_{1}$ & $p$ \\\hline\hline

  $0<\mu_{3}<a_{3}$ & $n_{1}<\mu_{1}\leqslant b_{5}$ & $c_{6}< p\leqslant c_{2}$ \\\cline{2-3}
                    & $b_{5}<\mu_{1}\leqslant b_{6}$ & $c_{6}<p\leqslant c_{7}$ \\\hline\hline

  $a_{3}\leqslant\mu_{3}\leqslant a_{1}$ & $n_{2}<\mu_{1}< b_{5}$ & $c_{6}< p\leqslant c_{2}$ \\\cline{2-3}
                                         & $b_{5}\leqslant\mu_{1}< b_{6}$ & $c_{6}<p\leqslant c_{7}$ \\\hline\hline

                       & $b_{5}\leqslant\mu_{1}<b_{6} $ & $c_{6}<p\leqslant c_{3} $ \\\cline{2-3}
  $a_{1}<\mu_{3}<1/3 $ & $b_{2}<\mu_{1}\leqslant b_{3} $ & $c_{6}<p< c_{3} $ \\\cline{2-3}
                       & $b_{3}<\mu_{1}<b_{7} $ & $c_{6}<p\leqslant c_{2} $ \\\hline\hline

                       & $b_{2}<\mu_{1}\leqslant b_{3} $ & $c_{6}<p< c_{3} $ \\\cline{2-3}
  $1/3<\mu_{3}<a_{2} $ & $b_{3}<\mu_{1}< b_{5} $ & $c_{6}<p\leqslant c_{2} $ \\\cline{2-3}
                       & $b_{5}\leqslant\mu_{1}<b_{8} $ & $c_{6}<p\leqslant c_{7} $ \\\hline\hline

  $\mu_{3}=a_{2}$      & $a_{1}<\mu_{1}< b_{9}$ & $c_{8}< p\leqslant c_{9}$ \\\cline{2-3}
                       & $b_{9}\leqslant\mu_{1}< b_{10}$ & $c_{8}<p\leqslant c_{10}$ \\\hline\hline

  $a_{2}<\mu_{3}<1$    & $n_{3}<\mu_{1}< b_{5}$ & $c_{6}< p\leqslant c_{2}$ \\\cline{2-3}
                       & $b_{5}\leqslant\mu_{1}< b_{8}$ & $c_{6}<p\leqslant c_{7}$ \\ \hline

\end{tabular}\\
\par
where
\par
$a_{3}=\frac{2641-1740\sqrt{2}}{7053}$,

$b_{5}=\frac{1}{2}[2-\mu_{3}-\sqrt{\mu_{3}(4-3\mu_{3})}]$,
$b_{6}=1-2\mu_{3}$,
$b_{7}=\frac{1}{2}(2-\mu_{3})$,
$b_{8}=\frac{1}{2}(1-\mu_{3})$,

$b_{9}=\frac{1}{58}(49-4\sqrt{2}-\sqrt{705+248\sqrt{2}})$,
$b_{10}=\frac{2}{29}(5-\sqrt{2})$,

$$
c_{6}=1-2\sqrt{\frac{1+6\mu_{1}^{2}-3\mu_{3}+6\mu_{3}^{2}-3\mu_{1}(1+\mu_{3})}{[\mu_{1}^{2}-2\mu_{1}(\mu_{3}-4)+\mu_{3}(8+\mu_{3})-32]^{2}}}
+\frac{22-3\mu_{1}-3\mu_{3}}{\mu_{1}^{2}-2\mu_{1}(\mu_{3}-4)+\mu_{3}(8+\mu_{3})-32}
$$
$c_{7}=\frac{\mu_{1}+\mu_{3}-1}{\mu_{1}+\mu_{3}-2}$,
$$
c_{8}=1+\frac{29(12\sqrt{2}-611+87\mu_{1})}{24711-1000\sqrt{2}-29\mu_{1}(214-8\sqrt{2}+29\mu_{1})}
$$
$$
-58\sqrt{\frac{736+84\sqrt{2}+174\mu_{1}(29\mu_{1}-19-2\sqrt{2})}{[1000\sqrt{2}-24711+1000\sqrt{2}+29\mu_{1}(214-8\sqrt{2}+29\mu_{1})]^{2}}}
$$
$c_{9}=1+\sqrt{\frac{29(9+4\sqrt{2})\mu_{1}}{[(9+4\sqrt{2})\mu_{1}-29]^{2}}}+\frac{29}{9+4\sqrt{2})\mu_{1}-29}$, $c_{10}=1+\frac{29}{4\sqrt{2}-49+29\mu_{1}}$. \\
Although $n_{1},n_{1},n_{1}$ exist but they have not simple algebraic form so we did't write them.

\newpage
\vspace{1cm}\setcounter{section}{0}


\newpage
{\bf Figure Captions}
\\\\
{\bf FIG. 1:} The entanglement detection in terms of a. All entangled states are detected for $a=[0,1]$.
\\\\
{\bf FIG. 2:} The entanglement detection region for the $d=4$ case in terms of parameters $\mu_{1}$ and $\mu_{3}$. All states in this region are $PPT$ entangled states.
\\\\
{\bf FIG. 3:} The entanglement detection region for the $d=5$ case in terms of parameters $\mu_{3}$ and $\mu_{4}$. All states in this region are $PPT$ entangled states.

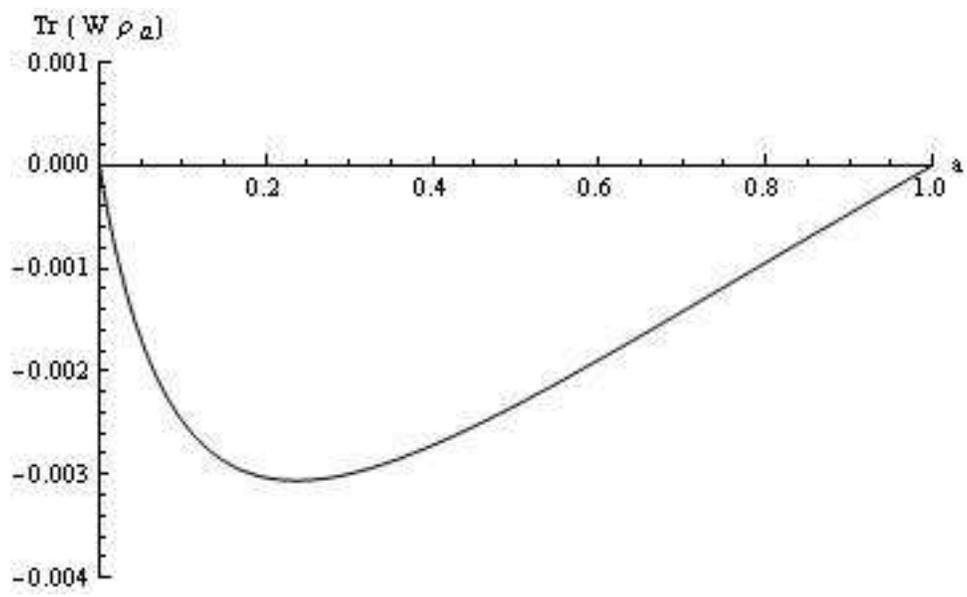
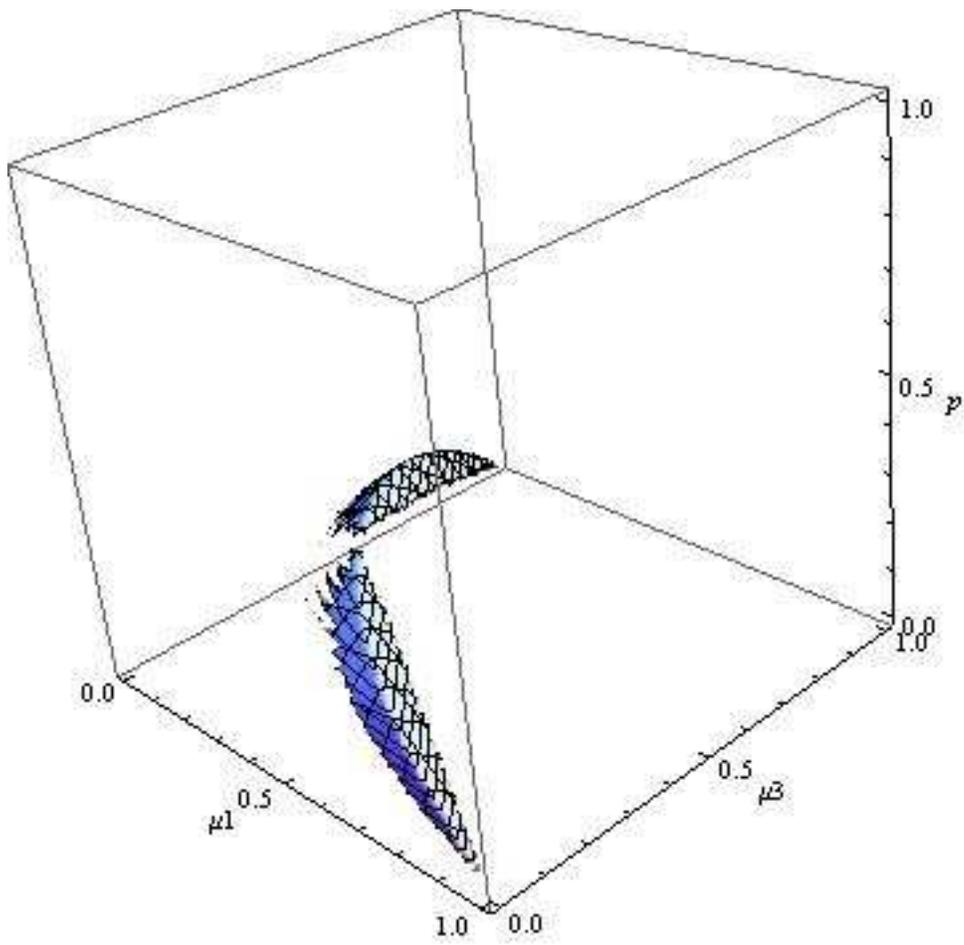
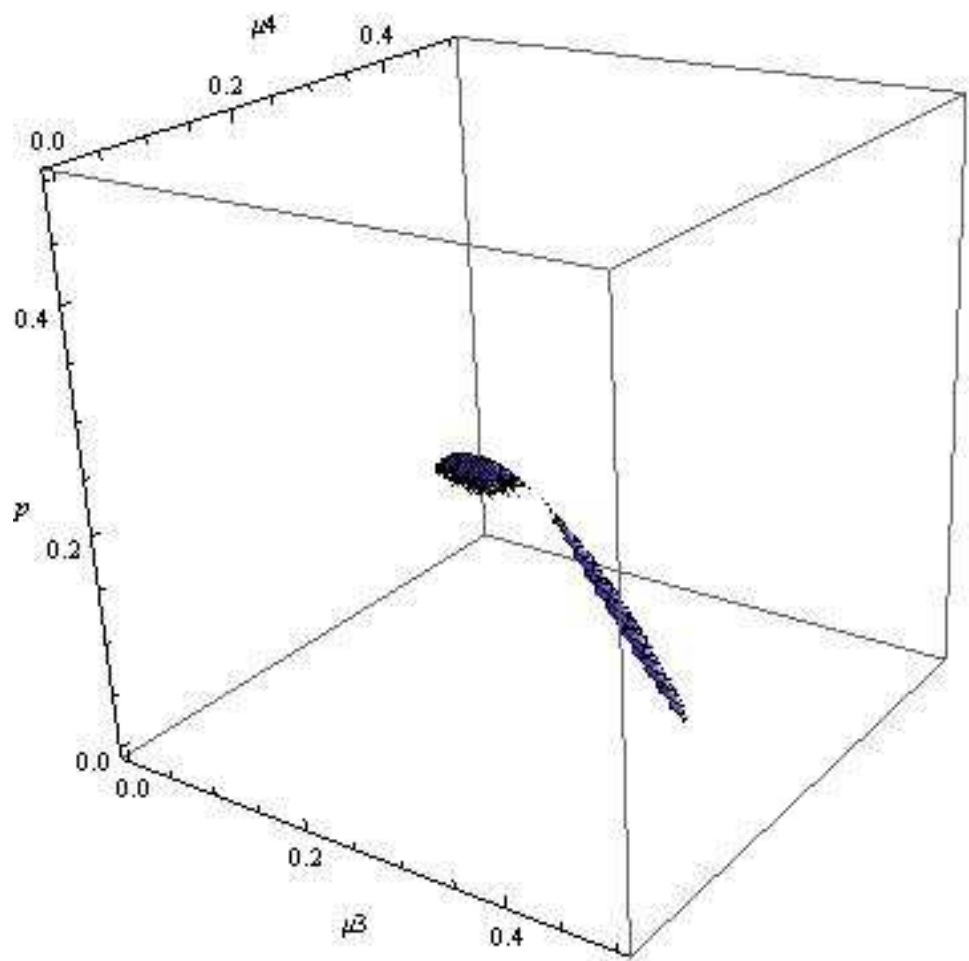
\end{document}